\documentclass[useAMS,usenatbib]{mn2e}
\usepackage{psfig}
\usepackage{epsfig}

\def\kms{\relax \ifmmode {\,\rm km\,s}^{-1}\else \,km\,s$^{-1}$\fi}

\def\mincir{\ \raise-2.truept\hbox{\rlap{\hbox{$\sim$}}\raise5.truept
    \hbox{$<$}\ }}
\def\magcir{\ \raise-2.truept\hbox{\rlap{\hbox{$\sim$}}\raise5.truept
    \hbox{$>$}\ }}
\def\gr{$^\circ$}
\def\sm{M$_\odot$}

\def\arcsec{\hbox{$^{\prime\prime}$}}
\def\nii{[N {\sc ii}]}
\def\sii{[S {\sc ii}]}

\def\oiii{[O {\sc iii}]}

\def\cliii{[Cl {\sc iii}]}
\def\ha{H$\alpha$}
\def\hb{H$\beta$}
\def\hg{H$\gamma$}
\def\hd{H$\delta$}

\def\chb{$c_{\rm H\beta}$}
\def\te{T$_e$}

\def\ne{N$_e$}
\def\pi{\rm {Paper~{\sc i}}}

\def\phe{He~1-1}
\def\pkj{KjPn~8}
\def\pic{IC~2149}
\def\pngc{NGC~7662}

\title[Physical properties and excitation of pairs of knots]
  {Low-ionization pairs of knots in planetary nebulae: physical properties and excitation}
\author[D. R. Gon\c calves et al.]
  {D. R. Gon\c calves$^1$, 
  A. Mampaso$^2$, R. L. M. Corradi$^{2,3}$
  and C. Quireza$^4$
\\
  $^1$UFRJ - Observat\'orio do Valongo, Ladeira Pedro Antonio 43, 20080-090, Rio
  de Janeiro, Brazil\\
  $^2$Instituto de Astrof\'\i sica de Canarias, E-38205 La Laguna, Tenerife,
  Spain\\
  $^3$Isaac Newton Group of Telescopes, Apartado de Correos 321, E-38700 Sta.
  Cruz de La Palma, Spain\\
  $^4$Observat\'orio Nacional, Rua General Jos\'e Cristino 77, 20921-400 Rio
  de Janeiro, Brazil}
\date{Released 2002 Xxxxx XX}

\pagerange{\pageref{firstpage}--\pageref{lastpage}} \pubyear{2002}

\def\LaTeX{L\kern-.36em\raise.3ex\hbox{a}\kern-.15em
    T\kern-.1667em\lower.7ex\hbox{E}\kern-.125emX}

\begin{document}

\label{firstpage}

\maketitle

\begin{abstract}
We obtained optical long-slit spectra of four planetary
nebulae (PNe) with low-ionization pair of knots, namely \phe, \pic,
\pkj\ and \pngc.

These data allow us to derive the physical parameters and excitation
of the pairs of knots, and those of higher ionization inner
components of the nebulae, separately.

Our results are as follows. 1) The electron temperatures of the knots
are within the range 9~500 to 14~500~K, similar to the
temperatures of the higher ionization rims/shells. 2) Typical knots'
densities are 500 to 2~000~cm$^{-3}$. 3) Empirical densities of the
inner rims/shells are higher than those of the pairs of knots, by up
to a factor of 10.  Theoretical predictions, at variance with the
empirical results, suggest that knots should be denser than the inner
regions, by at least a factor of 10. 4) Empirical and theoretical
density contrasts can be reconciled if we assume that at least 90\%
of the knots' gas is neutral (likely composed of dust and
molecules). 5) By using \citet{raga08} shock modeling and diagnostic diagrams
appropriated for spatially resolved PNe, we suggest that high-velocity
shocked knots traveling in the photoionized outer regions of PNe can
explain the emission of the pairs of knots analysed in this
paper.
\end{abstract}

\begin{keywords}
({\it ISM}): Planetary nebulae: individual: \phe, \pic, \pkj, \pngc. Planetary nebulae: structure.
\end{keywords}

\section{Introduction}

The small-scale low-ionization structures (LIS) of planetary nebulae were
classified in terms of their morphology and kinematics by Gon\c calves, Corradi
\& Mampaso (2001) as knots, jets and
jetlike\footnote{These are features whose morphology resemble highly supersonic jets, but that
share the expansion of the nebular components in which they are embedded.} systems,
either in pairs or isolated. From this, and subsequent works (e.g. \citealt{g2004}, and  references therein)
it has became apparent that: i) around 10\% of the Galactic PNe
are known to possess LIS; ii) they are indistinctly spread among all the PNe
morphological classes; iii) 50\% of these PNe have highly collimated,  high-velocity
jets, and/or  high-velocity pairs of knots (FLIERs; fast, low-ionization emission
regions, \citealt{bea1993});
and iv) most of them are mainly photoionized.

However, a number of key questions remain open: are there significant density contrasts
between LIS and the higher ionization main components (rims, attached shells and detached
haloes) of the nebula? What are the typical electron densities of the different
types of LIS?

We are involved in a long-term project for characterizing the
small-scale LIS of PNe since they can tell
us much about the formation and evolution of PNe. The main issues we
can access by studying the LIS are:

\begin{enumerate}
\item the collimation processes of LIS in regard to those processes responsible
for the shape of the PN itself, or the mass-loss processes during the AGB and
post-AGB phases;

\item the effect of the ionization front on the fossil AGB features; and

\item the role of disks and magnetic fields in shaping the highly
collimated outflows (jets) in PNe.
\end{enumerate}

Our motivation for the present analysis is to go a step further on
the study of the different LIS analyzing their densities. LIS can move through their 
environments either with high or low velocity. In the first case (the so-called FLIERS), 
radial velocities of $24-200$~\kms with respect to the PN main components are typically measured 
\citep{bea1993}, whereas in the second instance (the SLOWERs, slow moving low-ionization emission 
regions; see \citealt{p2000}) the knots 
do not show peculiar velocities and share the typical 30~\kms\ (for elliptical shells) 
to 100~\kms\ (for bipolar lobes) expansion velocities of their host PNe.
A straightforward question arises from this: 
how is the density contrast between the LIS and rims, shells or haloes
correlated with the different velocity regimes?

In this paper we analyse 4 PNe that contain pairs of knots, either
of high- or low-velocities.

We present our spectroscopic data and treatment, in Section~2. The line
fluxes obtained for as many as possible large- and small-scale
components of the four PNe, and the derived physical properties
for each of the components are shown in Sections 3 to 6. Finally,
Section 7, is dedicated to the discussion of the results in terms of
the match with the theoretical model predictions, and to our concluding remarks.

\section{Observation and Data Treatment}

\begin{table}
 \caption{Log of the INT+IDS observations}
\begin{tabular}{lcc}
\hline
PN Name and P.A. & Obs. Date & Exposure (s)\\
 &  & 3$\times$ \\
\hline
\phe\ - 315\gr\  & Aug 31 & 300, 2400	   \\
\pic\ - 70\gr\   & Sep 01 & 300, 900, 1800 \\
\pkj\ - 98\gr\   & Aug 30 & 1200           \\
\pkj\ - 120\gr\  & Aug 30 & 300, 1200      \\
\pngc\ - 175\gr\ & Sep 05 & 60,  300       \\
\pngc\ - 248\gr\ & Sep 04 & 30,  60,  300  \\
\hline
\end{tabular}
\label{LOG}
\end{table}

A 120 sec exposure  \ha\ + \nii\ (6568/95{\AA}) image of \phe\ (Figure~\ref{i_he11})
was retrieved from the IPHAS database (INT/WFC Photometric \ha\ Survey of the Northern
Galactic Plane: http://www.iphas.org).
We also retrieved from the Hubble Space Telescope (HST) Archive two WFPC2 images of
\pic: the V-band (filter 555W, 2 sec exp. time) and the \oiii\ (F502N, 100 sec) images,
obtained on 1995 (proposal 6119), see Figure~\ref{i_ic2149}. The \nii\ (F658N, 400 sec)
image of \pngc\ (Perinotto et
al. 2004; Figure~\ref{i_ngc7662}), was also retrieved from the HST archive, and was obtained
on 1996. The image of \pkj\ was obtained with the
2.5-m Nordic Optical Telescope (NOT) at the Observatorio del Roque de
los Muchachos (European Northern Observatory, La Palma, Spain), with
ALFOSC camera, in June 26, 2002. The image we show in
Figure~\ref{i_kjpn8} is the result of one exposure of 1 800~s through
the narrow-band filter \nii6584\AA.

Spectra of the 4 PNe were obtained on August 31 and September 1, 4 and
5 of 2001, at the 2.5 m Isaac Newton Telescope (INT) at the Observatorio del
Roque de los Muchachos using the Intermediate Dispersion Spectrograph (IDS). The 235
mm camera and the R300V grating were used, providing a spectral
coverage from 3650 to 7000~\AA\ with a spectral reciprocal dispersion
of 3.3~\AA~pixel$^{-1}$. The spatial scale of the instrument was
0\farcs70~pixel$^{-1}$, with the TEK5 CCD. Seeing varied from
0\farcs9 and 1\farcs1.
The slit width and length were 1\farcs5 and
4~\arcmin, respectively. These data were taken with the slits
positioned through the centre of the nebulae at the position angles
(P.A.), and with exposure times given in Table~1. Each of the
exposures was taken 3 times, as listed in the table. The longer
exposures were used to measure the fluxes of most emission
lines. However, for almost all the PNe, the
\oiii$\lambda\lambda$~4959,5007~{\AA}, the
\nii$\lambda\lambda$~6548,6583~{\AA} doublet as well as the \ha\ and even
the \hb\ lines were saturated at the brightest features in the longer
exposures, thus, the shorter exposures where used in these
cases. 
Differently of the way we took long-slit spectra of the
other PNe, in the case of \pkj\ the slit was centred on the position
of the central star, and then it was shifted in order to cover
each of the knots. Compared to the weather conditions during the
observation of the other three PNe, that of the \pkj\ was poor
(clouds appeared during the integration).

During the night, bias frames, twilight and tungsten flat-field exposures, wavelength calibrations,
and exposures of standard stars (BD +332642, Cyg OB2 No. 9, HD 19445, and BD +254655) were
obtained. Spectra were reduced following the IRAF instructions for long-slit spectra, being
bias-subtracted, flat-fielded, combined in order to improve the signal-to-noise ratio (S/N) and
eliminate cosmic rays, wavelength-calibrated, and sky-subtracted. Finally, they were flux-calibrated
using the above mentioned standard stars and the mean atmospheric extinction curve for La Palma.

\section{He 1-1}

\begin{figure}
\begin{center}
 \psfig{file=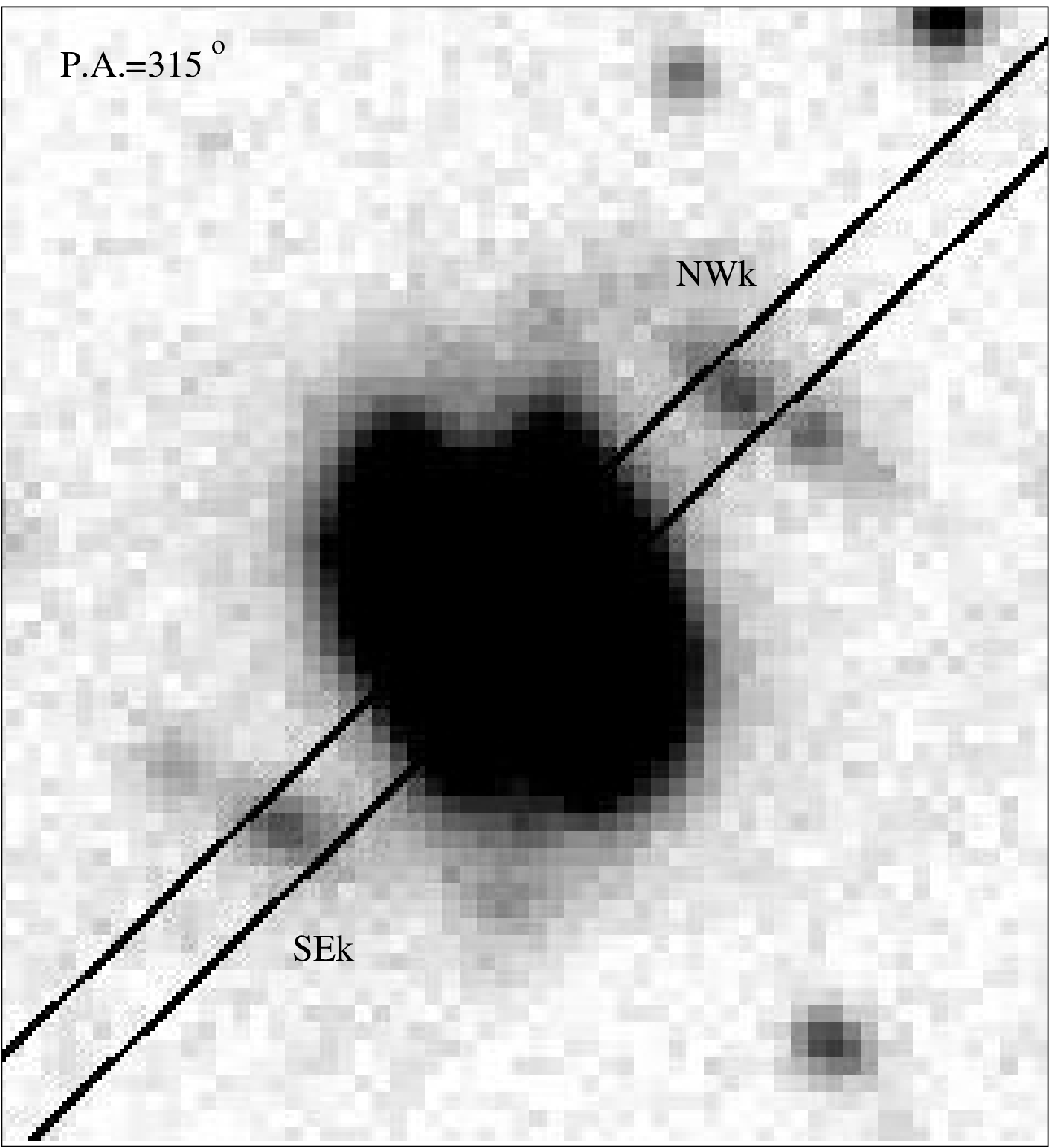,width=7.0truecm}
 \psfig{file=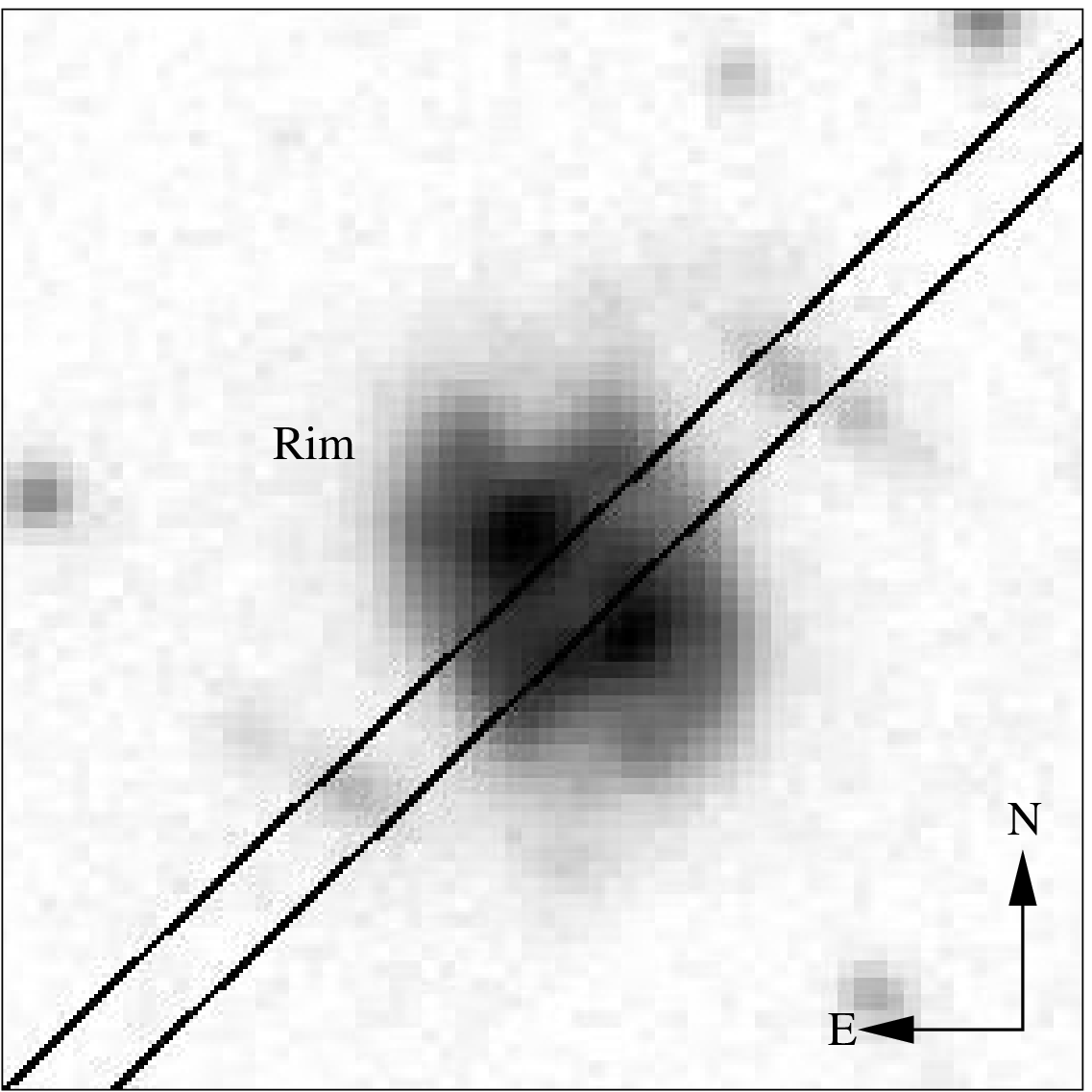,width=7.0truecm}
\end{center}
\caption{He~1-1: IPHAS \ha+\nii\ image. The 3 structures under analysis are
  indicated in the image.
  The size of the extraction windows in the spectra includes 4\farcs2 each
  side from the centre for the Rim, and extends from + 4\farcs9 to
  7\arcsec, and from -4\farcs9 to -7\farcs7 from the centre, for the
  SE and the NW knots respectively. The extraction window for the
  entire portion of the nebula covered by the slit, NEB, extends along 14\farcs7.}
 \label{i_he11}
\end{figure}

PN G055.3+02.7 (\phe) is a point-symmetric PN with a pair of
low-ionization knots.  See in Figure~\ref{i_he11} the extensions of
its bright Rim (or a combination of a rim and a shell), the large-scale
component that dominates the emission
of the nebula, and the SE and NW strings of knots, which correspond to
the small-scale faint structures seen in the \nii\ NOT image.

In the uppermost part of Table~2 
we list selected line
fluxes of the Rim, the knots and the entire portion of the nebula
covered by the slit. These include all the electron density and
temperature diagnostic emission lines covered by the IDS
spectra. Absolute \hb\ fluxes, F$_{{\rm H}\beta}$, integrated along
the slit for each nebular component are also given in Table~2.
The Balmer lines ratios (\ha/\hb, \hg/\hb\ and
\hd/\hb) were then used to derive the c$_{{\rm H}\beta}$, the
logarithmic ratio between observed and dereddened \hb\ fluxes. For the
derivation of \chb, we assumed $T_{\rm e}$ = 10$^4$~K and the
densities given by the \sii\ ratio of each region. Theoretical Balmer
line ratios from \citet{of2006} and the reddening law of
\citet{ccm1989} were used. The weighted average of \chb, per PN
component, is the one given in the table.  \chb\ is constant
along the slit, and the \chb=1.49$\pm0.16$ for the entire nebula is
lower than the value previously published by \citet{tea1992}. The
latter authors found a range of values for \chb, which goes from 1.9
to 2.2. Note, however, that their values are based only on the
\ha/\hb\ ratio. Fluxes were then dereddened using the derived
\chb. The dereddened fluxes, or intensities, are not shown in the
table. Instead, figures in the table correspond to the observed
fluxes.

\begin{table}
\centering
\begin{minipage}{90mm}
{\scriptsize
\caption{He~1-1: Observed line fluxes, flux errors, \chb, \ne\ and \te.}
\begin{tabular}{lcccc}
\hline
\hline
Line ID & \multicolumn{1}{c}{Rim} & \multicolumn{2}{c}{Knots} &
\multicolumn{1}{c}{NEB} \\
       & & SEk & NWk  & \\
\hline
\noalign{\smallskip}
{}[S{\sc ii}] 4072.0     & 6.232   & -      & -      & 6.464 \\
H$\delta$ 4101.8         & 15.57   & -      & -      & 15.97 \\
H$\gamma$ 4340.5         & 29.69   & -      & -      & 29.13 \\
{}[O{\sc iii}] 4363.2    & 11.36   & 16.89  & 14.61  & 13.04 \\
H$\beta$ 4861.3          & 100.0   & 100.0  & 100.0  & 100.0 \\
{}[O{\sc iii}] 4958.9    & 543.4   & 587.7  & 601.7  & 566.8 \\
{}[O{\sc iii}] 5006.86   & 1715.   & 1773.  & 1808.  & 1774. \\
{}[CL{\sc iii}] 5517.7   & 3.321   & 22.56  & 15.57  & 4.152 \\
{}[CL{\sc iii}] 5537.9   & 3.041   & 18.23  & 13.04  & 3.640 \\
{}[N{\sc ii}] 5754.6     & 11.90   & 23.46  & 24.87  & 13.83 \\
{}[N{\sc ii}] 6548.0     & 319.2   & 480.6  & 413.9  & 335.1 \\
H$\alpha$ 6562.8         & 993.9   & 977.6  & 981.2  & 1021. \\
{}[N{\sc ii}] 6583.4     & 1034.   & 1436.  & 1321.  & 1075. \\
{}[S{\sc ii}] 6716.5     & 129.3   & 114.4  & 124.4  & 132.3 \\
{}[S{\sc ii}] 6730.8     & 169.6   & 114.7  & 139.2  & 172.1 \\
\\
F$_{{\rm H}\beta}$ $^a$  & 21.62         & 0.4565        & 0.6247        & 22.37 \\
c$_{{\rm H}\beta}$       & 1.46$\pm$0.20 & 1.44$\pm$0.32 & 1.45$\pm$0.33 & 1.49$\pm$0.16 \\
\\
 & \multicolumn{4}{c}{Percentage errors in line fluxes} \\
\\
(0.01--0.05)\hb\  &35  &52   &60   &31 \\
(0.05--0.15)\hb\  &15  &25   &27   &13 \\
(0.15--0.30)\hb\  &11  &20   &20   &11 \\
(0.30--2.0)\hb\   &10  &17   &16.5 &8.5\\
(2.0--5.0)\hb\    &9.5 &15   &16   &6.5\\
(5.0--10.0)\hb\   &9   &14.5 &16   &6.5\\
$>$ 10\hb\         &9   &14.5 &16   &6.5\\
\\
& \multicolumn{4}{c}{Electron Densities (cm$^{-3}$) and Temperatures (K)} \\
\multicolumn{4}{l}{}\\
\ne\sii   	& 1600$\pm$230   & 600$\pm$140    & 900$\pm$210    & 1550$\pm$190   \\
\ne\cliii 	& 1700$\pm$830   & 800$\pm$230    & 1000$\pm$330   & 1350$\pm$580   \\
\te\oiii        & 12500$\pm$2060 & 14400$\pm$3320 & 13450$\pm$4110 & 13050$\pm$1580 \\
\te\nii         & 10800$\pm$1800 & 12900$\pm$2980 & 14000$\pm$3320 & 11400$\pm$1590 \\
\te\sii         &  8750$\pm$1450 & -              & -              &  9300$\pm$1350 \\
\hline
\hline
\multicolumn{5}{l}{{\bf $^a$} In units of 10$^{-15}$erg cm$^{-2}$ s$^{-1}$} \\
\end{tabular}
}
\end{minipage}
\label{t_he11}
\end{table}

The middle part of Table~2 
shows the errors on the
fluxes. They were calculated taking into account the statistical errors
in the measurements, as well as systematic errors of the
flux calibrations, background determination, and sky
subtraction.

In the lowermost part of Table~2 
electron densities and temperatures are presented. At least two different
measures of \te\ and \ne\ are available for each PN component. They
represent regions of low (\ne\sii, \te\nii\ and \te\sii) and medium
excitation (\ne\cliii\ and \te\oiii).

The density of the nebula can be compared with \citet{skas1992}.  They
find \ne\sii=1.19$\times$10$^3$~cm$^{-3}$, in moderate agreement with
ours. Figures for \te\ and \ne\ of the knots were previously published
only in a proceeding contribution by \citet{benitez02}: 11~300~K for
the temperature of the NW knot, and densities of 300~cm$^{-3}$ and
600~cm$^{-3}$, for the SE and NW knots respectively.  These values
hardly agree with ours. Note that they did not derive, but assumed
the 10$^4$~K of the SE knot.

\begin{figure*}
\psfig{file=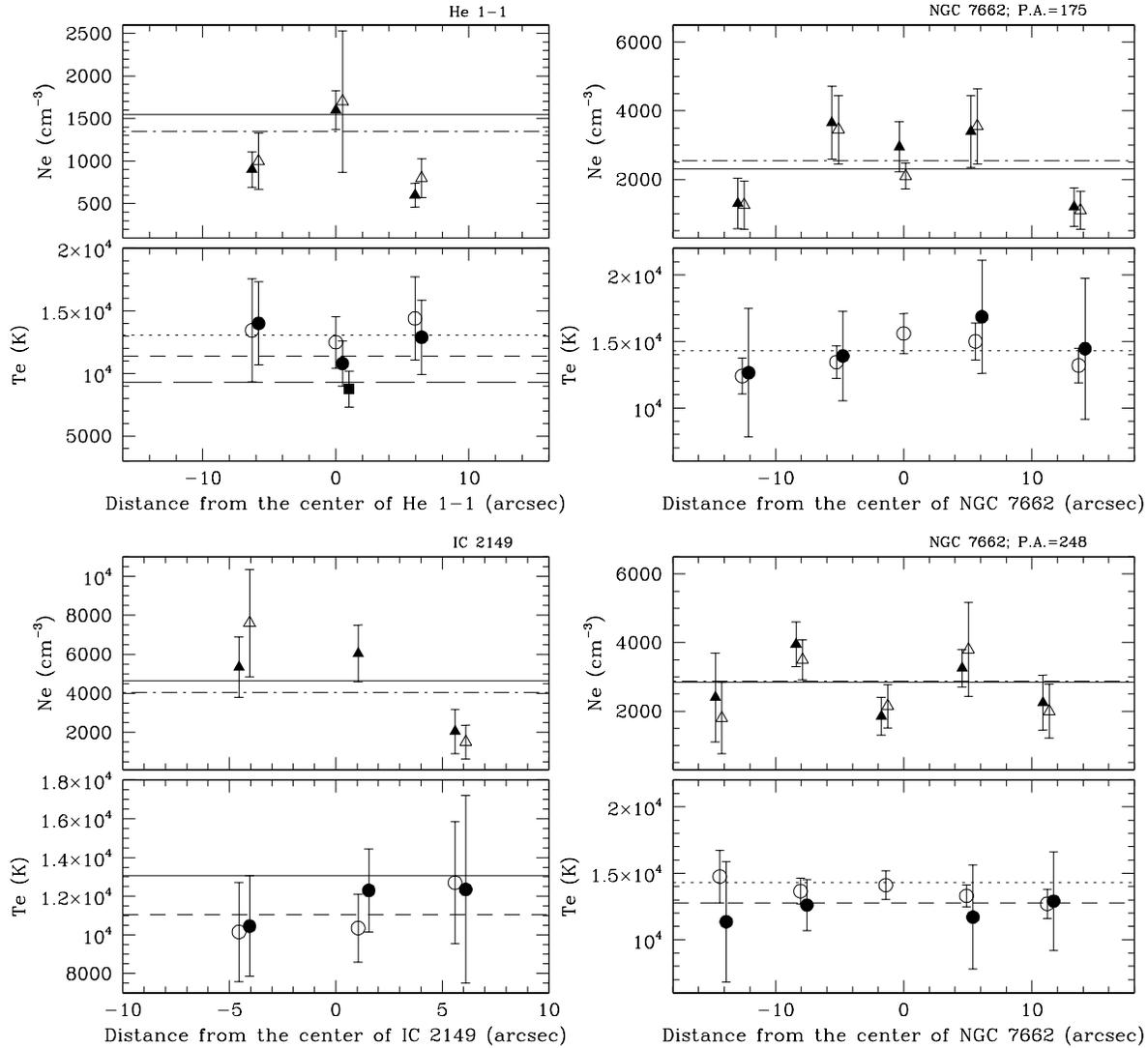,width=16.0truecm}
\caption{Electron densities and temperatures as a function of the distance to the centre of
the nebula. The different structures under analysis are shown as symbols that correspond
to the centre of each structure, as defined in the images (Figures~1, 3, 4, and 5). Filled
(\ne\sii) and open (\ne\cliii) triangles represent densities, while filled (\te\nii) and open
(\te\oiii) circles, as well as the filled box (\te\sii) represent electron
temperatures. Symbols
are plotted slightly displaced in distance in order to avoid overlapping. Horizontal lines
represent  densities and temperatures averaged along the slit (NEB): continuous line for \ne\sii, large-small
dashed line for \ne\cliii, small dashed line for \te\oiii, medium dashed line for \te\nii,
and large dashed line for \te\sii.
TOP LEFT: \phe\ -  from left to right the NW knot, the Rim and the SE knot, respectively,
as in  Figure~\ref{i_he11}.
BOTTOM LEFT: \pic\ - From left to right, the NE knot,
the Core and the SW knot as in Figure~\ref{i_ic2149}.
TOP RIGHT: \pngc\ -
 P.A.=175$^{\circ}$; five zones of the Inner rim and Outer shell, as shown in Figure~\ref{i_ngc7662} and Table~5.
 BOTTOM RIGHT:\pngc\ -
 P.A.=248$^{\circ}$; further five zones of the Inner rim and FLIERs, following Figure~\ref{i_ngc7662}
 and Table~5.
  }
\label{td_mosaic}
\end{figure*}

Figure~\ref{td_mosaic} (top left panel) and Table ~2 shows that both
density measurements in each region of the nebula are equal
within the errors, whereas \ne\ of the Rim is higher than that of the
SE and NW knots, by factors of $\sim$2.7 and $\sim$1.8,
respectively.
There is a good agreement between the two temperature
diagnostics obtained for the three PN components.

Figure~\ref{td_mosaic} shows that \te\ at the centre could be somewhat lower
than at the knots, but the variation is within the errors.

Summarising, the pair of knots in \phe\ are substantially less dense
than the Rim, whereas no strong evidence exist for significant temperature
variations from the Rim to the knots.

\section{IC 2149}

\begin{figure}
\psfig{file=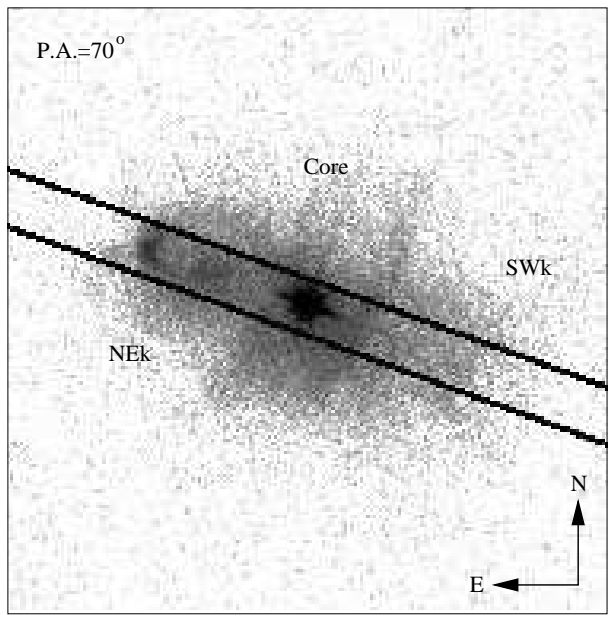,width=8.0truecm}
 \caption{\pic: HST 555W (V) image. Since this filter is centred in \oiii, only the Core
 is clearly identified in this image. The low-ionization NEk and its fainter SWk counterpart
 are hard to see in this image, but are easily identified in the \nii\ image in Fig.1 of
 \citet{vel2002}.
 The spectroscopic sizes of the Core, NE  and SW knots are, respectively: 2\farcs1
 each side from the centre; - 2\farcs8 to - 8\farcs4; and + 2\farcs8 to 6\farcs3.
 The entire PN, NEB, is 14\farcs7 in size.}
 \label{i_ic2149}
\end{figure}

\begin{table}
\centering
\begin{minipage}{90mm}
{\scriptsize
\caption{IC 2149: Observed line fluxes, flux errors, \chb, \ne\ and \te}
\begin{tabular}{lcccc}
\hline
\hline
Line ID & \multicolumn{1}{c}{Core} & \multicolumn{2}{c}{Knots} &
\multicolumn{1}{c}{NEB}\\
       & & NEk  &  SWk  &  \\
\hline
\noalign{\smallskip}
{}[S{\sc ii}] 4072.0     & -      & 5.743  & 5.625 & 3.011 \\
H$\delta$ 4101.8         & 15.75  & 26.99  & 36.27 & 24.14 \\
H$\gamma$ 4340.5         & 40.46  & 43.41  & 61.99 & 46.12 \\
{}[O{\sc iii}] 4363.2    & 3.838  & 2.457  & 5.333 & 3.850 \\
H$\beta$ 4861.3          & 100.0  & 100.0  & 100.0 & 100.0 \\
{}[O{\sc iii}] 4958.9    & 180.1  & 124.5  & 140.7 & 150.9 \\
{}[O{\sc iii}] 5006.86   & 540.2  & 378.4  & 408.4 & 454.0 \\
{}[CL{\sc iii}] 5517.7   & -	  & 0.367  & 0.545 & 0.386 \\
{}[CL{\sc iii}] 5537.9   & - 	  & 0.516  & 0.481 & 0.440 \\
{}[N{\sc ii}] 5754.6     & 1.140  & 1.633  & 1.227 & 1.299 \\
{}[N{\sc ii}] 6548.0     & 15.03  & 30.27  & 18.55 & 22.06 \\
H$\alpha$ 6562.8         & 320.6  & 321.3  & 277.8 & 315.1 \\
{}[N{\sc ii}] 6583.4     & 43.07  & 91.54  & 51.81 & 65.23 \\
{}[S{\sc ii}] 6716.5     & 1.288  & 3.201  & 2.418 & 2.261 \\
{}[S{\sc ii}] 6730.8     & 2.312  & 5.607  & 3.385 & 3.860 \\
\\
F$_{{\rm H}\beta}$ $^a$  & 11.17     & 12.78          & 4.549 & 31.76 \\
c$_{{\rm H}\beta}$       & 0.07$^b$  & 0.14$\pm$0.015 & 0.00  & 0.12$\pm$0.008 \\
\\
 & \multicolumn{4}{c}{Percentage errors in line fluxes} \\
\\
(0.01--0.05)\hb\  &17   &24  &39  &10  \\
(0.05--0.15)\hb\  &10   &17  &24  &5.5 \\
(0.15--0.30)\hb\  &7    &11  &16  &4.5 \\
(0.30--2.0)\hb\   &6.5  &8   &10  &4.5 \\
(2.0--5.0)\hb\    &5.5  &7   &8.5 &4.5 \\
(5.0--10.0)\hb\   &4.5  &6.5 &7   &4   \\
$>$ 10 \hb\       &4.5  &6   &6.5 &4   \\
\\
& \multicolumn{4}{c}{Electron Densities (cm$^{-3}$) and Temperatures (K)} \\
\multicolumn{4}{l}{}\\
\ne\sii           & 6050$\pm$1450  & 5350$\pm$1550  & 2050$\pm$1130 & 4650$\pm$660 \\
\ne\cliii         & -              & 7600$\pm$2750  & 1500$\pm$870  & 4000$\pm$680  \\
\te\oiii          & 10350$\pm$1760 & 10150$\pm$2560 &12700$\pm$3150 & 11050$\pm$1170 \\
\te\nii           & 12300$\pm$2150 & 10450$\pm$2600 &12350$\pm$4850 & 11000$\pm$1170 \\
\hline
\hline
\multicolumn{5}{l}{{\bf $^a$} In units of 10$^{-14}$erg cm$^{-2}$ s$^{-1}$} \\
\multicolumn{5}{l}{{\bf $^b$} This c$_{{\rm H}\beta}$ values is the average between the knots' values; see text.} \\
\end{tabular}
}
\end{minipage}
\label{t_ic2149}
\end{table}

PN G166.1+10.4 has an apparent shape that does not easily compare
with the classification bins (round, elliptical, bipolar or
quadrupolar, irregular and point-symmetric) usually adopted for
PNe. Kinematic modelling suggests that it is a bipolar PN
(\citealt{vel2002}; \citealt{fhl1994}; \citealt{zk1998}).  As shown in Figure~\ref{i_ic2149},
\pic\ is composed by a bright higher
ionization emission zone, the Core, and by a pair of lower excitation
knots, the North-Eastern one being much brighter than its
counterpart to the South-West. In fact only
the two former structures can be identified from
Figure~\ref{i_ic2149}. Please refer to Fig.1 of \citet{vel2002}, in
which not only a higher excitation image is show, but also the lower
excitation \nii\ and \sii\ are presented.  In their \nii\ image the SW LIS is also clearly seen.

Table~3 
contains the spectroscopic results for \pic, obtained from the INT+IDS
spectra at P.A.=70\gr\ (observed and absolute \hb\ fluxes, flux errors, \chb,
\ne\ and \te) for the Core and pair of LIS of the nebula.

The \ha/\hb\ ratio was used to obtain the \chb\ of the NEk, SWk
and NEB regions. Owing to the presence of Balmer line absorption in
the central zones \citep{vel2002}, the \chb\ of the Core was adopted
as the average from the knots' values. The value measured for the
nebula \chb=0.12$\pm0.01$ is in agreement with other published values
(\citealt{fhl1994}; \citealt{ciardullo1999}), but it is much smaller
than the value (0.41) obtained by \citet{vel2002} for the Core and the knots.

The gas physical conditions we derived to this PN (Table~3 and left bottom panel
of Fig.~2) compare relatively well with those derived by
\citet{vel2002} for the North-Eastern knot and for the NEB. The
exception is the \ne\sii, for which they obtained values of 10~000 and
9~600~cm$^{-3}$ while we found 5~350$\pm$1~550 and
4~650$\pm$660~cm$^{-3}$, for the NEk and NEB, respectively. We cannot
understand the reason of this discrepancies, since they do not depend
on the very different \chb\ applied to correct the fluxes. On the
other hand, the \te\ that would be affected by our different choices
in terms of \chb, are in reasonable agreement with theirs (see
\citealt{vel2002} Table~1).  If only NEB is concerned, our results
also agree with the results obtained by \citet{fhl1994}.

As [Cl~III] $\lambda\lambda$5517,5537\AA\ were not detected, only
\ne\sii\ could be derived for the Core.  Figure~\ref{td_mosaic} shows
that the two density diagnostics for each of the knots are such that
SWk (\ne\sii=2~050$\pm$1~130) is significantly less dense than NEk,
whose density is similar to that in the Core (\ne\sii=5~350$\pm$1~550 and
6~050$\pm$1~450, respectively).

In spite of an apparent increase of \te\oiii\ and \te\nii\ from the
North-Eastern to the South-Western side of the nebula, variations
are within the errors.


\section{\pkj}

PN G112.5-00.1 (\pkj) is a huge PN (size $\sim$ 14~\arcmin $\times$
4~\arcmin; L\'opez, V\'azquez \& Rodr\'\i guez 1995) with multiple
polar ejections.  A number of papers  deal with the
characteristics of this PN, including: i) narrow-band imaging of its
many structures in the different scales, \citet{lvr1995}; ii) proper
motions of its latest bipolar ejections, \citet{M1997}; iii) kinematics
of the large bipolar lobes and of the pair of knots,
\citet{lmbr1997}; and iv) temperature, density and chemical abundances of
the inner nebula, V\'azquez, Kingsburgh \& L\'opez (1998). At much
smaller scales, a ring that seems to collimate the many structures
(with its CO and H$_2$ counterpart) has been detected giving support to
the idea that \pkj\ had two PNe-like events in its history
(\citet{lea2000}). However, \ne\ and \te\ of its two pairs of
low-ionization knots were not properly measured so far, due to their
extreme faintness.

In our spectra only one of the knots of each pair could be measured with
a good S/N. At variance with the other PNe in our sample, no analysis of
the internal or entire nebula will be given, because of the huge extension
of the PN. Our results for the two SE knots of
\pkj\ are given in Table~4. 

\begin{figure}
 \psfig{file=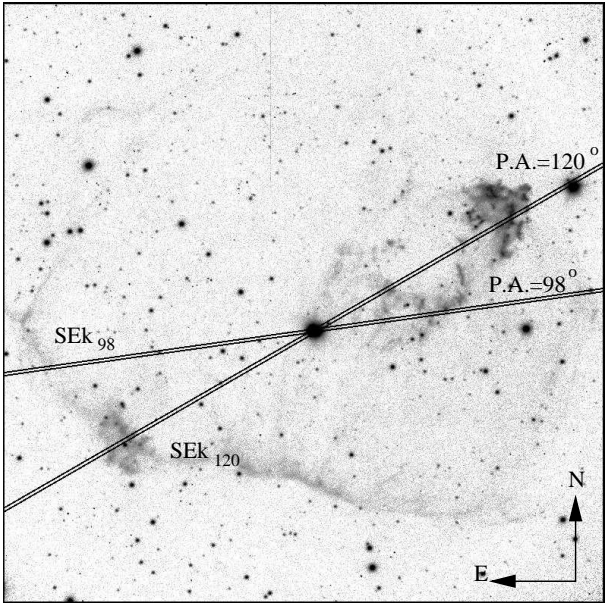,width=8.0truecm}
 \caption{\pkj: NOT \nii\ image. The two P.A. of the structures under analysis
 are indicated in the image.
 The spectroscopic sizes of the two knots, SEk$_{98}$ is 14\arcsec\ and
 that of  SEk$_{120}$ is 10\farcs5.}
 \label{i_kjpn8}
\end{figure}

\begin{table}
\centering
\begin{minipage}{90mm}
{\scriptsize
\caption{KjPn8: Observed line fluxes, flux errors, \chb, \ne\ and \te}
\begin{tabular}{lcc}
\hline
\hline
Line ID & \multicolumn{2}{c}{South Eastern Knots} \\
       & SEk$_{120}$  & SEk$_{98}$ \\
\hline
\noalign{\smallskip}
{}[S{\sc ii}] 4072.0      & 10.78 & 8.869 \\
H$\delta$ 4101.8          & 23.35 & 20.69 \\
H$\gamma$ 4340.5          & 41.41 & 42.54 \\
{}[O{\sc iii}] 4363.2     & 1.149 & 2.466 \\
H$\beta$ 4861.3           & 100.0 & 100.0 \\
{}[O{\sc iii}] 4958.9     & 100.5 & 101.8 \\
{}[O{\sc iii}] 5006.86    & 307.9 & 307.4 \\
{}[CL{\sc iii}] 5517.7    & -	  & 3.000 \\
{}[CL{\sc iii}] 5537.9    & -	  & 2.280 \\
{}[N{\sc ii}] 5754.6      & 9.351 & 12.95 \\
{}[N{\sc ii}] 6548.0      & 392.8 & 398.0 \\
H$\alpha$ 6562.8          & 438.8 & 453.3 \\
{}[N{\sc ii}] 6583.4      & 1191. & 1207. \\
{}[S{\sc ii}] 6716.5      & 99.79 & 98.28 \\
{}[S{\sc ii}] 6730.8      & 100.4 & 98.19 \\
\\
F$_{{\rm H}\beta}$ $^a$   & 2.468          & 1.221         \\
c$_{{\rm H}\beta}$        & 0.48$\pm$0.07  & 0.52$\pm$0.10 \\
\\
 & \multicolumn{2}{c}{Percentage errors in line fluxes} \\
\\
(0.01--0.05)\hb\  & 30.5  & 42.5 \\
(0.05--0.15)\hb\  & 22.5. & 34.5 \\
(0.15--0.30)\hb\  & 15.5  & 27.5 \\
(0.30--2.0)\hb\   & 10.5  & 16   \\
(2.0--5.0)\hb\    &  8.5  & 10   \\
(5.0--10.0)\hb\   &  6.5  & 8.5  \\
$>$ 10\hb\        &  5    & 8.0  \\
\\
& \multicolumn{2}{c}{Electron Densities (cm$^{-3}$) and Temperatures (K)} \\
\multicolumn{2}{l}{}\\
\ne\sii        & 600$\pm$90     & 600$\pm$140  \\
\ne\cliii      & -	        & 450$\pm$270	\\
\te\oiii       & 9100$\pm$2960  & 11550$\pm$4950 \\
\te\nii        & 8500$\pm$1950  &  9550$\pm$3340 \\
\te\sii        & 9150$\pm$1250  &  8400$\pm$3010 \\
\hline
\hline
\multicolumn{3}{l}{{\bf $^a$} In units of 10$^{-14}$erg cm$^{-2}$ s$^{-1}$}\\
\end{tabular}}
\end{minipage}

\label{t_kjpn8}
\end{table}

When comparing our \chb\ values with those in the literature, particular care should be taken because they usually refer to different nebular regions, and reddening may vary along this large nebula.
\citet{vkl1998} measured very different \chb\ values for the three knots which they  labelled A1, A3 and B1: \chb\ $=$ 1.49, 0.96 and 0.56, respectively. Our \chb\ for
SEk$_{120}$ (their A1) is therefore 3 times smaller than
theirs, whereas for SEk$_{98}$ (their knot B1) both determinations agree very well, \chb=0.56 and 0.52$\pm0.10$,
respectively. The reason for the discrepant \chb\ in SEk$_{120}$ is not clear at present. As for the
density, they reported \ne\sii=100~cm$^{-3}$ for both SEk$_{98}$ and SEk$_{120}$, while we
derived substantially larger values \ne\sii=600$\pm$140 (\cliii = 450$\pm$270) and 600$\pm$90 for
SEk$_{98}$ and SEk$_{120}$, respectively.  The electron temperature in
the knots is measured in this paper for the first time.

Summarising, the densities and temperatures of the two knots are the same
within the errors.

\section{NGC~7662}

\begin{figure}
 \begin{center}
 \psfig{file=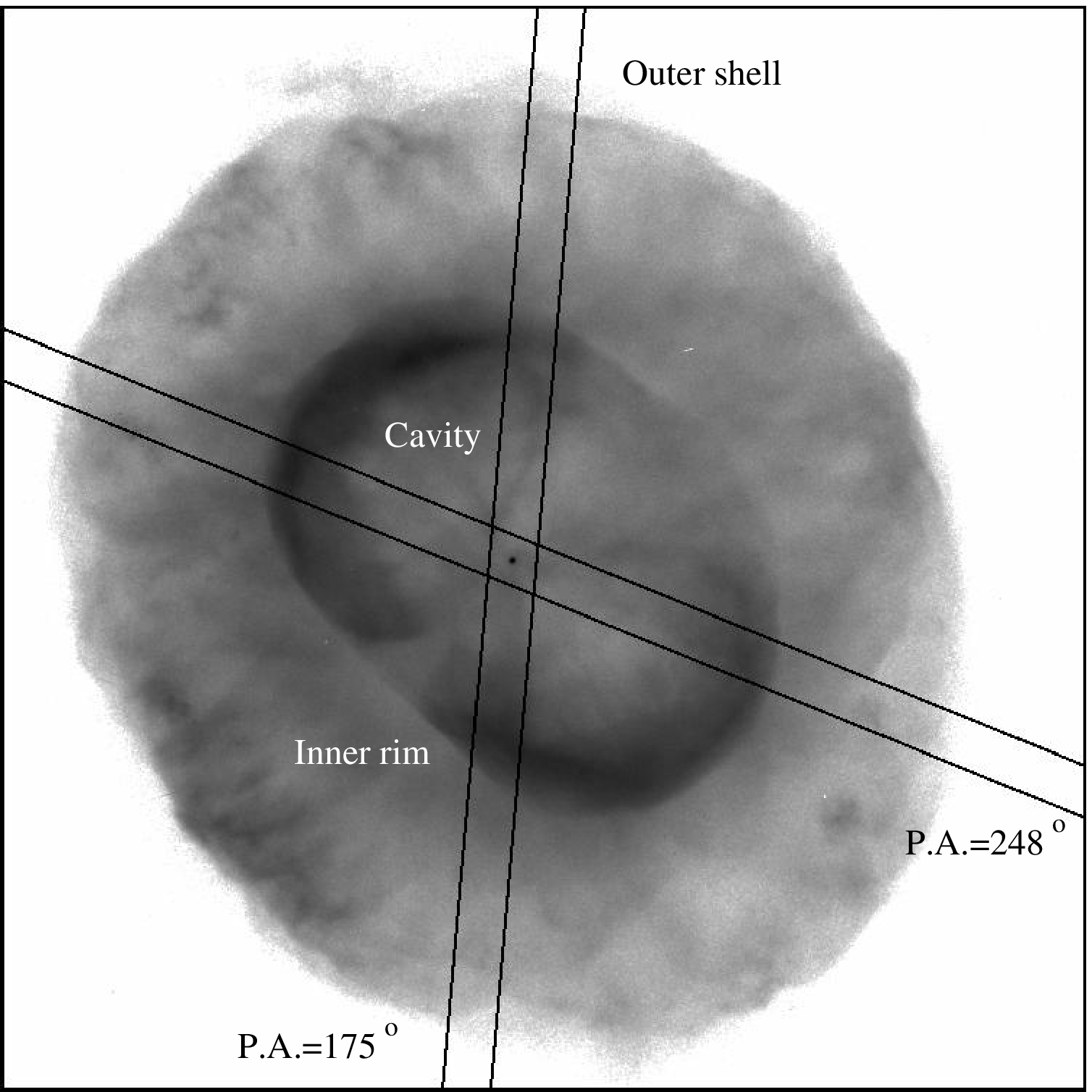,width=7.0truecm}
 \psfig{file=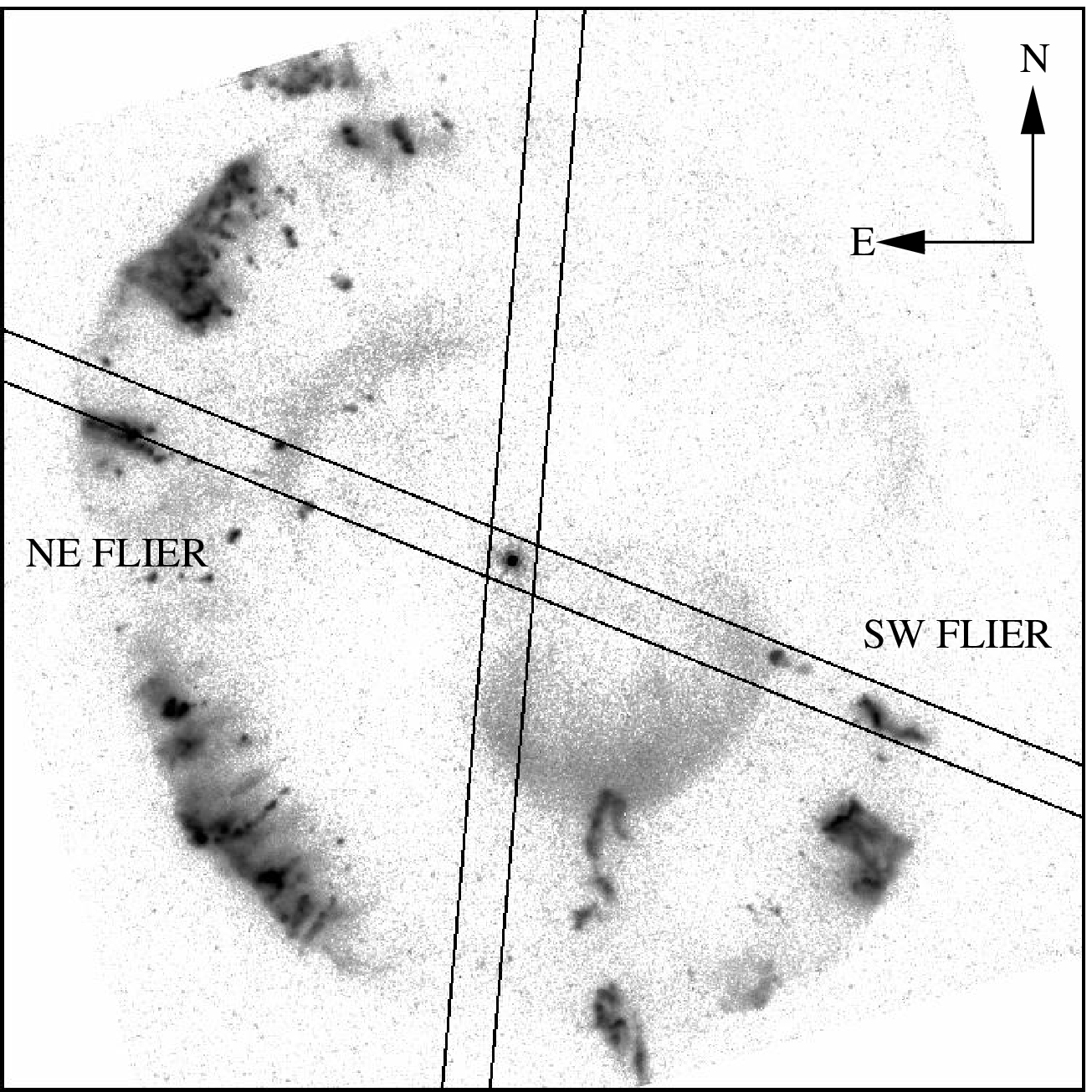,width=7.0truecm}
 \end{center}
 \caption{\pngc: HST archive images. {\it Top}: \oiii. {\it Bottom}: \nii. In both images the
 box corresponds to 34$\times$34~arcsec$^2$.
 The size of the extraction windows in the spectra are as follows.
  {\it P.A.=175$^{\circ}$ (inner rim + outer FLIER)}: -14\farcs7 to -10\farcs5, SE outer shell;
                                                 -8\farcs4 to -2\farcs8, SE inner rim;
                                                 -1\farcs75 to +1\farcs75, Cavity + Star;
                                                 +2\farcs8 to +8\farcs4, NW inner rim;
						+12\farcs6 to +14\farcs7, NW outer shell.
  {\it  P.A.=248$^{\circ}$ (inner rim + outer shell)}:  -14\farcs7 to -11\farcs2, SW FLIER;
                                                 -9\farcs8 to -2\farcs8, SW inner rim;
                                                 -2\farcs1 to +2\farcs1, Cavity + Star;
                                                 +3\farcs5 to +9\farcs8, NE inner rim;
						+10\farcs5 to +15\farcs4, NE FLIERS. }
\label{i_ngc7662}
\end{figure}

This bright nebula (PN G106.5-17.6) consist of a
central cavity surrounded by three concentric shells, the inner Rim,
the outer shell \citep{gjc2004}, and a halo \citep{corradi04}.  In addition, on
smaller scales it has a number of LIS. Most of them appear, in
projection, distributed at the outer edge of the outer shell.  All
these structures are easily identified in Figure~\ref{i_ngc7662}, and
are named as in \citet{bea1998} and
\citet{pea2004}. This figure also identify the structures that we
analyze in this paper. \citet{bpi1987} studied the kinematics of the
different nebular components, from which the acronyms FLIERs and
SLOWERs were later introduced.

\citet{pea2004} studied this PN with unprecedented spatial resolution
using the STIS spectrograph on the HST. However, their spectra were taken
through two slits (called POS1 and POS2, see their Figure~1), that in terms
of LIS, are restricted to
only one side of the nebula. We have used much longer slits that cover all
the structures along both sides of the nebula, in each direction at P.A.
175$^{\circ}$ and 248$^{\circ}$. This allows us to compare the densities
and temperatures of any given symmetrical
pair of LIS. The slit at P.A. 248$^{\circ}$ contains a pair of
low-ionization filaments (the serpentine-like filaments described in
\citealt{bea1998}). The slit at P.A. 175$^{\circ}$ does not contain any LIS, and is
used to compare the physical parameters along the line joining the
FLIERs with those in a direction free of micro structures. In this
way, the present work can be seen as complementary to the analysis of
\citet{pea2004}.

In Table~5 
we report the derived values of \chb\, that are based on the
average weighted \chb\ given by the \ha, \hg, and \hd\ to \hb\ ratios.
These values are all between 0.13 and 0.21 for the 5 regions under analysis in
each position angle. These \chb\ agree very well with those previously published:
0.18 - 0.22 \citep{tea1992}; 0.1 \citep{ha1997}; 0.2 \citep{pea2004};
and 0.18 \citet{zea2004}.

This table 
gives the observed line fluxes, densities and temperatures of the
10 structures we measured along P.A. 175$^{\circ}$ and 248$^{\circ}$. These
parameters are also presented in Figure~\ref{td_mosaic}.
The average \te\oiii\ derived from the
emission integrated along both P.A. are the same within errors
(\te\oiii=14
300$\pm$700~K and 13 400$\pm$680~K, respectively).  In fact these
values also agree with the \te\nii=12 750$\pm$1 530~K derived along
the P.A.=248$^{\circ}$.
The \te\oiii\ we obtained here are more in agreement with the
results by \citet{b1986} -- whose \oiii\ temperatures in different
regions of the nebula (see his Table~3) varies from 11 200~K to 13
800~K -- and \citet{zea2004} (13~300~K), than with the somewhat lower
values of \citet{pea2004}.

As for the \sii\ and \cliii\ densities of the integrated emission we
found 2~300$\pm$440~cm$^{-3}$ and 2~550$\pm$490~cm$^{-3}$ for
P.A.=175$^{\circ}$ and 2~850$\pm$450~cm$^{-3}$ and
2~850$\pm$460~cm$^{-3}$ for P.A.=248$^{\circ}$. \ne\sii\ results very
closely reproduce those found by a
\citet[][\ne\sii=2\,884~cm$^{-3}$]{wea2004}.  The same authors give a
value of 1~862~cm$^{-3}$ for the \cliii\ density, but errors are not
reported in their work.

Focusing on the densities and temperatures of the inner Rim and
outer shell with respect to those of the FLIERs
(pair of filaments at P.A.=248$^{\circ}$), it is clear from
Figure~\ref{td_mosaic}, that: i) temperatures \te\oiii (where
errors are smaller) are slightly higher at the inner
Rim than in the outer knots (P.A.=175$^{\circ}$), but this is
just at the 2-$\sigma$ level of the adopted errors; ii) densities of the Rim
agree very well one to another in both position angles; iii) the Rim is
roughly a factor of 2.8 denser than the outer shell
(P.A.=175$^{\circ}$); iv) the Rim (inner shell) along P.A.=248$^{\circ}$,
is a factor of $\sim$~1.7 denser than its outer region and the pair
of FLIERs.

Summarising, \citet{pea2004} previously observed the LIS (FLIERs and SLOWERS)
of this PN at different directions than in the present work. Densities for the
FLIERs in both works are in good agreement: \ne=2 200~cm$^{-3}$ for the two
FLIERs studied by \citet{pea2004}, and 2 100~cm$^{-3}$ in our work. \citet{pea2004}
did find significantly different densities for FLIERs and SLOWERs, being the
latter approximately 1.3 times denser than the former.

Our analysis shows that FLIERS of \pngc\ are less dense than the inner regions
of the nebula, but they are at the same time much denser than the PN outer shell
in which they are located.

\begin{table*}
\centering
\begin{minipage}{120mm}
{\scriptsize
\caption{NGC 7662 Observed line fluxes, flux errors, \chb, \ne\ and \te.
P.A.=175$^{\circ}$ (Rim and Outer shell) and P.A.=248$^{\circ}$ (Rim + FLIERs)}
\begin{tabular}{llcccccc}
\hline
\hline
P.A.=175$^{\circ}$ &  Line ID    & SE outer Shell  & SE inner Rim & Cavity + Star & NW inner Rim & NW outer Shell & NEB\\
\hline
&\multicolumn{7}{l}{}\\
&H$\delta$ 4101.8       &  30.81 &  27.31 &  28.38 & 29.48 &  24.25 & 28.99 \\
&H$\gamma$ 4340.5       &  23.79 &  23.99 &  22.00 & 23.99 &  24.60 & 23.99 \\
&{}[O{\sc iii}] 4363.2  &  17.17 &  15.66 &  19.78 & 18.46 &  19.12 & 17.82 \\
&H$\beta$ 4861.3        &  100.0 &  100.0 &  100.0 & 100.0 &  100.0 & 100.0 \\
&{}[O{\sc iii}] 4958.9  &  504.4 &  368.0 &  343.2 & 338.7 &  481.7 & 372.1 \\
&{}[O{\sc iii}] 5006.86 &  1533. &  1105. &  1018. & 1018. &  1434. & 1114. \\
&{}[CL{\sc iii}] 5517.7 &  0.761 &  0.317 &  1.756 & 0.345 &  0.398 & 0.424 \\
&{}[CL{\sc iii}] 5537.9 &  0.652 &  0.344 &  1.669 & 0.378 &  0.475 & 0.423 \\
&{}[N{\sc ii}] 5754.6   &  0.169 &  0.087 &  0.469 & 0.099 &  0.152 & -     \\
&H$\alpha$ 6562.8       &  331.5 &  324.9 &  334.9 & 326.7 &  334.9 & 339.9 \\
&{}[N{\sc ii}] 6583.4   &  7.496 &  2.981 &  2.502 & 2.540 &  5.191 & 3.115 \\
&{}[S{\sc ii}] 6716.5   &  1.164 &  0.366 &  0.242 & 0.249 &  0.916 & 0.371 \\
&{}[S{\sc ii}] 6730.8   &  1.428 &  0.593 &  0.373 & 0.396 &  1.107 & 0.536 \\
\\
&F$_{{\rm H}\beta}$$^a$ & 3.164          & 36.90          & 14.15          & 30.94          & 2.138          & 104.4 \\
&c$_{{\rm H}\beta}$     & 0.18$\pm$0.017 & 0.16$\pm$0.012 & 0.20$\pm$0.016 & 0.17$\pm$0.015 & 0.18$\pm$0.017 & 0.21$\pm$0.02\\
\\
& & \multicolumn{6}{c}{Percentage errors in line fluxes} \\
\\
&(0.01--0.05)\hb\ &35  &15.5 &12.5 &16.5 &30  &8.5 \\
&(0.05--0.15)\hb\ &16  &10.5 &10   &9.5  &11  &4.5 \\
&(0.15--0.30)\hb\ &10  &8.5  &9    &8.5  &9   &4   \\
&(0.30--2.0)\hb\  &7   &5.5  &7.5  &6	 &7   &3.5 \\
&(2.0--5.0)\hb\   &5.5 &4.5  &5    &5.5  &5.5 &3.5 \\
&(5.0--10.0)\hb\  &5   &4    &4.5  &5	 &5   &3.5 \\
&$>$ 10\hb\       &4.5 &4    &4    &4.5  &5   &3.5 \\
& & \multicolumn{6}{c}{Electronic Densities (cm$^{-3}$) and Temperatures (K)} \\
\\
&\ne\sii   & 1300$\pm$735   & 3650$\pm$1060  & 2950$\pm$730   & 3400$\pm$1050  & 1200$\pm$560	& 2300$\pm$440  \\
&\ne\cliii & 1250$\pm$700   & 3450$\pm$1000  & 2100$\pm$370   & 3550$\pm$1100  & 1100$\pm$550	& 2550$\pm$490  \\
&\te\oiii  & 12400$\pm$1350 & 13450$\pm$1220 & 15600$\pm$1500 & 15000$\pm$1390 & 13200$\pm$1300 & 14300$\pm$700 \\
&\te\nii   & 12650$\pm$4830 & 13900$\pm$3350 & -	      & 16850$\pm$4250 & 14450$\pm$5300 & -		\\
\\
\hline
P.A.=248$^{\circ}$ & Line ID   & SW outer   & SW inner & Cavity + Star & NE inner  & NE outer & NEB \\
                   &           & FLIER      & Rim      &               & Rim       & FLIER    &   \\
\hline
\multicolumn{7}{l}{}\\
&{}[S{\sc ii}] 4068.6   & 1.420 & 0.885 & 0.814 & 0.920 & 1.263 & 1.232 \\
&{}[S{\sc ii}] 4076.4   & 0.530 & 0.469 & 0.966 & 0.430 & 0.317 & 0.259 \\
&{}[S{\sc ii}] 4072.0   & 1.923 & 0.954 & 2.127 & 1.131 & 1.381 & 1.804 \\
&H$\delta$ 4101.8       & 24.39 & 24.98 & 25.08 & 23.69 & 23.93 & 23.99 \\
&H$\gamma$ 4340.5       & 43.96 & 45.22 & 45.07 & 44.17 & 43.99 & 45.29 \\
&{}[O{\sc iii}] 4363.2  & 21.83 & 17.27 & 16.43 & 16.53 & 18.89 & 16.98 \\
&H$\beta$ 4861.3        & 100.0 & 100.0 & 100.0 & 100.0 & 100.0 & 100.0 \\
&{}[O{\sc iii}] 4958.9  & 552.7 & 387.6 & 339.8 & 398.1 & 517.8 & 397.1 \\
&{}[O{\sc iii}] 5006.86 & 1706. & 1184. & 1040. & 1214. & 1570. & 1213. \\
&{}[CL{\sc iii}] 5517.7 & 0.637 & 0.395 & 0.408 & 0.389 & 0.606 & 0.410 \\
&{}[CL{\sc iii}] 5537.9 & 0.585 & 0.431 & 0.390 & 0.437 & 0.570 & 0.422 \\
&{}[N{\sc ii}] 5754.6   & 0.585 & 0.088 & -	& 0.089 & 0.552 & 0.155 \\
&H$\alpha$ 6562.8       & 343.8 & 331.9 & 321.6 & 331.3 & 333.8 & 321.1 \\
&{}[N{\sc ii}] 6583.4   & 31.14 & 3.551 & 2.734 & 4.289 & 22.88 & 6.321 \\
&{}[S{\sc ii}] 6716.5   & 2.630 & 0.454 & 0.311 & 0.431 & 2.141 & 0.611 \\
&{}[S{\sc ii}] 6730.8   & 3.843 & 0.751 & 0.423 & 0.679 & 3.064 & 0.932 \\
\\
&F$_{{\rm H}\beta}$$^a$ & 3.580          & 33.59          & 22.23          & 41.22          & 8.732          & 117.9\\
&c$_{{\rm H}\beta}$     & 0.21$\pm$0.025 & 0.17$\pm$0.011 & 0.14$\pm$0.011 & 0.17$\pm$0.011 & 0.19$\pm$0.015 & 0.14$\pm$0.008\\
\\
&  & \multicolumn{6}{c}{Percentage errors in line fluxes} \\
\\
&(0.01--0.05)\hb\  & 38.5 &10.5 &   18 & 9.5 & 25.5 & 8.5 \\
&(0.05--0.15)\hb\  & 19.5 & 8.5 &  9.5 &   7 &   14 & 5.5 \\
&(0.15--0.30)\hb\  &   13 & 6.5 &  7.0 & 5.5 &    8 & 4.5 \\
&(0.30--2.0)\hb\   &  8.5 & 4.5 &  5.0 & 4.5 &    6 & 3.5 \\
&(2.0--5.0)\hb\    &  6.5 & 4.0 &  4.5 & 4.0 &  4.5 & 3.0 \\
&(5.0--10.0)\hb\   &  5.0 & 3.5 &  4.0 & 3.5 &    4 & 3.0 \\
&$>$ 10\hb\        &  4.5 & 3.5 &  3.5 & 3.5 &    4 & 3.0 \\
&  & \multicolumn{6}{c}{Electronic Densities (cm$^{-3}$) and Temperatures (K)} \\
\\
&\ne\sii   & 2400$\pm$1300  & 3950$\pm$650   & 1850$\pm$550   & 3250$\pm$550   & 2250$\pm$800	& 2850$\pm$450   \\
&\ne\cliii & 1800$\pm$1040  & 3500$\pm$580   & 2150$\pm$630   & 3800$\pm$1370  & 2000$\pm$790	& 2850$\pm$460   \\
&\te\oiii  & 14750$\pm$1980 & 13650$\pm$970  & 14100$\pm$1070 & 13300$\pm$820  & 12700$\pm$1100 & 13400$\pm$680  \\
&\te\nii   & 11350$\pm$4540 & 12600$\pm$1920 & -	      & 11700$\pm$3920 & 12900$\pm$3720 & 12750$\pm$1530 \\
\hline
\hline
&\multicolumn{7}{l}{{\bf $^a$} In units of 10$^{-13}$erg cm$^{-2}$ s$^{-1}$. Note that
[N{\sc ii}] 6548\AA\ was not measured for any of the structures. Thus, \te\nii}\\
&\multicolumn{7}{l}{were calculated assuming the theoretical relation between
[N{\sc ii}] 6583\AA\ and [N{\sc ii}] 6548\AA.}\\
\end{tabular}
}
\end{minipage}
\label{t_n7662}
\end{table*}

\section{Discussion}

\subsection{Temperature of the knots}

We have estimated individual temperatures \te\oiii\ and \te\nii\ for the different regions
of the four nebulae studied in this paper. As it is well known, temperature errors are
always important and they range in our case typically from around 1000 K for \te\oiii\ in
the brighter zones to more than 4000 K for \te\nii\ in the fainter regions. Therefore, a
sensible comparison of temperatures for the different components of each PN is precluded,
and only rather general conclusions can be extracted: 1) temperatures are those of normal
photoionized PNe, varying from 9~500 to 14~500~K typically; 2) \te\oiii\ agrees  within
the errors with \te\nii\ for every PN component measured. 3) No relevant differences between
\te\oiii\ at the LIS and at the higher excitation PN structures are found.

\subsection{Knots density contrasts}

The empirical density contrasts between the knots and the inner nebular regions,
as described in this paper, are such that knots are typically 1 to 3 times {\it less dense}
than the rims or cores of these PNe (\phe, \pic\ and \pngc): in no one of the LIS measured
here the electronic density is definitely larger than at the main PN structures. This
unexpected result is nevertheless also found in the full sample of PN with LIS.
Table~4 of \citet{gcm2001} lists 26 PNe with pairs of knots similar to those in the
present work. We have searched in the literature for density measures both
at the inner rims (or shells) of each PNe and at their pairs of knots, finding
eight\footnote{NGC~4634 \citep{guerrero08}; Hb~4 \citep{hajian1997}; IC~4593
\citep{corradi97}; NGC~7009 \citep{gea2003}; K~3-35 \citep{miranda00}; NGC~6826
\citep{bea1994}; NGC~2440 \citep{cuesta00}; K~1-2 \citep{exter03}.} PNe (in addition
to the four nebulae here studied) with adequate data. In all but one PN (NGC~6826)
the densities of the knots are {\it equal or lower} than those of the inner rims
or shells, by factors between 1 and 10.

\begin{figure*}
\psfig{file=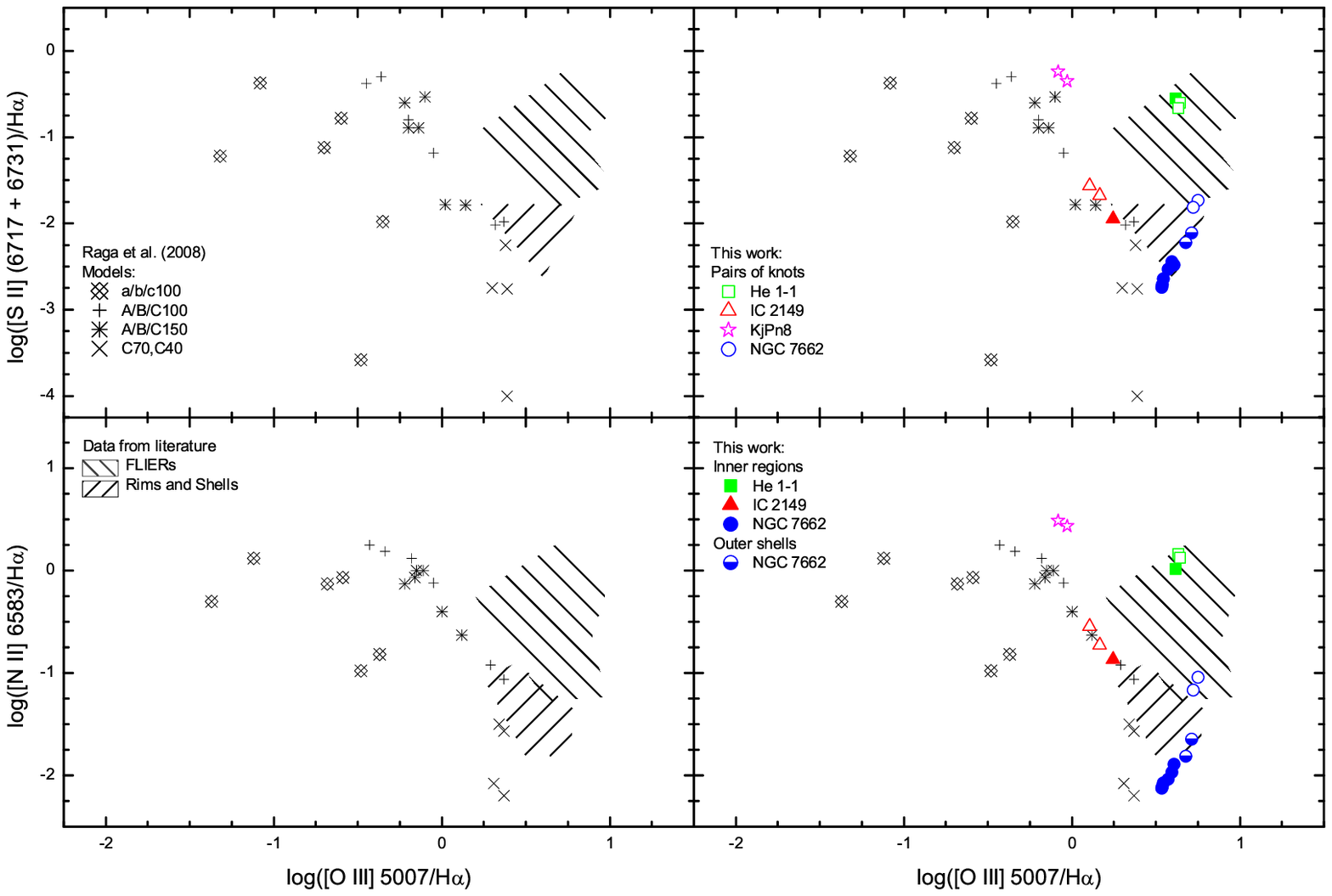}
\caption{Diagnostic diagrams for spatially resolved PNe.
Left panels are an adaptation of the \citet{raga08} results, whereas right panels superpose to these results the data for our sample.
The \citet{raga08} simulations for high-velocity shocked knots that travel through a photoionized region of a PN involve four families of models, as indicated in the labels of the upper left panel. Coded in their names, the numbers represent the knot's velocity, from 40 to 150 ~\kms whereas letters from `a' to `C' indicate different stellar luminosities and temperatures which are kept constant in each model while the knot's distance to the ionizing source decreases. Therefore, model `a100' has the lowest, and `C100' the highest, photoionization rate, while both assume a velocity of 100 ~\kms for the knots.
A compilation of empirical literature data performed by \citet{raga08} is also sketched in the four panels as regions filled with diagonal lines.
In the two right panels, different regions of the PNe of our sample are added: empty symbols correspond to the pairs of knots, filled symbols to the nebular inner regions (Rim, Core, Cavity + Star), and half-filled circles correspond to the outer shells.}
\label{raga}
\end{figure*}

This result is not consistent with the theoretical expectations. The formation of PN major
structures and pairs of jets/knots has been modeled in the past using a broad variety of
processes and scenarios (see, for instance, \citet{gcm2001} and \citet{bf2002}). Recent
models can be grouped into three main families. i) Magnetohydrodynamic disk wind models,
based on the scenario of accretion disks formed by binary interactions
(\citealt{fb2004}, \citealt{blackman2001}). ii) Interacting AGB and post-AGB wind
models, in which the post-AGB wind is driven exclusively by the magnetic pressure of a
single star (Garc\'\i a-Segura, L\'opez \& Franco 2005, \citealt{garcia2008}).
iii) Models that explore the effect of axisymmetrical light (low density) jets that,
for a short period of time, expand into a spherical AGB wind \citep{akashi2008}. Some of
the above models are particularly tailored to reproduce the main morphological and kinematic
properties of a few well observed PNe: He~3-401, M2-9 and He~2-90 \citep{g-segura2005}; Mz~3
and the pre-PN M1-92 \citep{akashi2008}. Nevertheless, and despite the different assumptions
of the different models, all of them are able to account for the formation of PNe with jets.

Model predictions state that, after the eventual shut off of the jets, {\it dense} knots would
form embedded in the outer nebular components. Moreover, both the knots present at the tips of
the existing jets and those left over when jets disappear are predicted to be denser than the
inner PN components by large factors ranging from 10 up to 1000 (typically 10) depending on the
model assumptions.

Therefore, there is a clear disagreement between observed and modeled pairs of knots: the former
are found to be typically 10 times less dense, while the latter are expected to be typically 10
times denser, than the inner rims and shells of their parent PN.

A straightforward possibility to reconcile the empirical and theoretical density contrasts is
keeping in mind that empirical densities, as those measured here, correspond to only the ionized
fraction of the PN gas. Models, on the other hand, usually predict the formation of knots accounting
for its morphology, kinematics and total gas content, but do not detail the ionization status of each
component nor their evolution with time. The strong discrepancy between models and observations found
here (a typical factor of 100) can therefore be explained if 90\% or more of the low-ionization knots'
material is neutral. It is worth mentioning here the case of one prominent cometary globule of the
Helix nebula --which contains the best known low-ionization knots of PNe. The molecular matter (H$_2$ + CO;
\citealt{huggins02}) and the dust content \citep{meaburn92} of this knot amount to $\sim$2$\times$10$^{-5}$\sm,
whereas \citet{ODell96} estimated that the typical mass of the photoionized gas of a globule in Helix is
$\sim$10$^{-9}$\sm.

In another well-studied system of knots, NGC~7009, previous modeling by \citet{gecac06} succeeded in reproducing the observed ionization structure and chemical abundances under the assumption that the total gas density of the knots is equal to the one empirically derived from the \sii\ lines \citep{gea2003}, therefore, not accounting for any neutral matter. This apparently contradicts the present results suggesting that the knots might be mostly neutral. However, the effect of the density assumed for the knots in the modeling by \citet{gecac06} is partly balanced by its geometry, i.e., by its shape and size. As the shape was fixed based on the knots' appearance in the HST images, the size was allowed to vary until it fits the observations (i.e. the optical line ratios and electron density). For that reason \citet{gecac06} were able to reproduce the observations of NGC~7009 without invoking need for neutral matter. It would be very interesting to attempt 3D photoionization modeling as in \citet{gecac06} exploring high density regimes for the knots as suggested in this paper.

\subsection{Main excitation of the pairs of knots}

\citet{raga08} have recently simulated the evolution of high-density (10$^3$~cm$^{-3}$),
high-velocity (100 to 150~\kms) knots that travel away from a photoionizing source
crossing a uniform lower density (10$^2$~cm$^{-3}$) environment. They explored a range of
initial conditions for the shocked knots (in terms of photoinization rate), and qualitatively
reproduced the emission-line ratios of several FLIERs, rims and shells observed in pre-PNe and
PNe. At the end of the model evolution, 400~yr, the number densities for the knots are
10$^3$ to 10$^4$~cm$^{-3}$, broadly in agreement with measured \ne\ of knots in NGC~7009 and IC~4634.
These simulations were indeed intended to reproduce the line ratios of the FLIERs of these two PNe.
No predictions are given for the properties of the other PN components (rims, shells and haloes).
 In Figure~6 we show two of the diagnostic diagrams from \citet{raga08} comparing
empirical and theoretical line ratios for FLIERs, as well as showing a compilation of empirical
data for rims and shells of some PNe.

Fig.~6 shows, first, that most of the inner regions (rims and shells) of the PNe in our
sample occupy the same zone as the rims and shells of other PNe in the literature. Second, the
same is
true for the zone occupied by the pairs of knots
in \phe\ and \pngc\ (open boxes and circles) with respect to the zones of the FLIERs
sample in \citet{raga08}. Third, knots in the other two nebulae, \pic\ and \pkj\
(open triangles and stars), are displaced from the empirical zone of FLIERs in \citet{raga08}.
Their locations are consistent with the results of simulations with the highest initial
velocity of the knots (150~\kms, model A/B/C/150), and indicate that a mixing of shock
and photo ionization is playing a role in the measured emission-line ratios. In fact,
the highly supersonic velocity (radial velocity of 226\kms; \citealt{lmbr1997}) measured
in \pkj\  for the SE knot at P.A.=98$^{\circ}$ could explain this agreement. Expansion
velocities of \pic's knots are more moderate ($\sim$ 47\kms; \citealt{vel2002}).

\citet{raga08} show that models with low photoionization rates yield line ratios typical
of shock-excited knots, whereas in models with higher photoionization rate, photoionized
emission dominates the spectrum of the high-velocity, high-density knots. This strengthen
the idea that it is evolution (PN age; see \citealt{g2004},
\citealt{raga08}; \citealt{vii2009}) and photoionization rate (\citealt{d1997};
\citealt{raga08}), rather than the highly supersonic velocities of the knots, the key
parameters to separate the shock-excited from the photoionized structures in planetary
nebulae.

\section{Conclusions}

We have derived the physical parameters of low-ionization pairs of knots,
and of higher ionization PN main components  for a small sample of nebulae (\phe, \pic,
\pkj\ and \pngc). We have shown that electron temperatures at the knots are comparable
to the temperatures of the main nebular components, rims and shells, whereas densities
in the latter structures are significantly higher than in the pairs of knots, by up to
a factor of 2. Typical knots' densities in the studied PNe are 500 to 2~000~cm$^{-3}$.

We have argued that optical line ratios are not appropriate to constrain the total gas density, as they only account for the ionized matter of the knots, whereas none of the available models properly explore the knots ionization fraction, but the total (neutral plus ionized) density structure. A direct comparison of our results, plus compiled literature data, with the theoretical density contrasts expected from model predictions, yields a clear discrepancy that amount, typically, up to a factor of 100. To reconcile models and observations we suggest that an important fraction of the gas at the knots --90\%-- should be neutral. 
The results presented here can help in constraining more realistic models for jets and
knots in PNe, by providing empirical physical parameters of the ionized gas.

We have analyzed the location of the knots of our four PNe in diagnostic diagrams studied
by \citet{raga08} concluding that the observed line emission ratios are compatible with
shocked knots traveling in the photoionized outer regions of their host PNe.

\section*{Acknowledgments}
We would like to thank the anonymous referee of this manuscript, for his suggestions. 
D.R.G. thanks the Brazilian agency FAPERJ (E-26/110.107/2008) for
its partial support. A.M., and R.L.M.C. acknowledge funding from the Spanish Ministry of
Science AYA2007-66804 grant. This paper makes use of data obtained
at the 2.5~m Isaac Newton Telescope (INT: IDS and IPHAS), operated by the Isaac Newton Group;
and at the 2.5~m Nordic Optical Telescope (NOT: ALFOSC), operated by NOTSA. Both are telescopes
of the European Northern Observatory  on the island of La Palma in the Spanish Observatorio del Roque
de los Muchachos of the Instituto de Astrof\'{i}sica de Canarias. We also use NASA/ESA
Hubble Space Telescope data, obtained at the Space Telescope Science Institute,
which is operated by AURA for NASA under contract NAS5-26555.



\label{lastpage}
\end{document}